\documentclass[sigconf]{acmart}
\usepackage{ifthen}      
\usepackage[normalem]{ulem}
\usepackage{url}
\usepackage{environ}
\usepackage{colortbl}
\usepackage{booktabs}
\usepackage{makecell}
\usepackage{CJKutf8}
\usepackage{multirow}

\newcommand{\ipstart}[1]{\vspace{1mm} 
\noindent{\textbf{\textit{#1.}}}}

\newcommand{\hashtag}[1]{\texttt{\##1}}
\definecolor{tableheader}{HTML}{EFEFEF}
\definecolor{tablegrayline}{HTML}{e0e0e0}

\newboolean{clean}
\setboolean{clean}{true} 

\newboolean{finalclean}
\setboolean{finalclean}{true} 

\newcommand{\revised}[1]{\ifthenelse{\boolean{clean}}{#1}{\textcolor{blue}{#1}}}
\newenvironment{revisedenv}{%
    \begingroup
    \ifthenelse{\boolean{clean}}{}{%
        \color{blue}%
    }%
}{%
    \endgroup
}

\newcommand{\finalrevised}[1]{%
  \ifthenelse{\boolean{finalclean}}{#1}{%
    \textcolor{purple}{#1}%
  }%
}

\newcommand{\deleted}[1]{%
  \ifthenelse{\boolean{clean}}{}{%
    \textcolor{gray}{}%
  }%
}

\newcommand{\deletedsubsection}[1]{%
  \ifthenelse{\boolean{clean}}{}{%
    \subsection{\textcolor{gray}{[Deleted] #1}}%
  }%
}

\newcommand{\deletedsubsubsection}[1]{%
  \ifthenelse{\boolean{clean}}{}{%
    \subsubsection{\textcolor{gray}{[Deleted] #1}}%
  }%
}

\newcommand{\todo}[1]{%
  \ifthenelse{\boolean{clean}}{}{%
    {}
  }%
}

\copyrightyear{2026}
\acmYear{2026}
\setcopyright{cc}
\setcctype{by}
\acmConference[CHI '26]{Proceedings of the 2026 CHI Conference on Human Factors in Computing Systems}{April 13--17, 2026}{Barcelona, Spain}
\acmBooktitle{Proceedings of the 2026 CHI Conference on Human Factors in Computing Systems (CHI '26), April 13--17, 2026, Barcelona, Spain}
\acmPrice{}
\acmDOI{10.1145/3772318.3790766}
\acmISBN{979-8-4007-2278-3/2026/04}

\begin{document}
\begin{CJK}{UTF8}{mj}



\title[Constructing Everyday Well-Being]{Constructing Everyday Well-Being: Insights from God-Saeng (God生) for Personal Informatics}

\author{Inhwa Song}
\orcid{0009-0000-2325-663X}
\affiliation{%
  \department{Department of Computer Science}
  \institution{Princeton University}
  \city{New Jersey}
  \country{USA}
}
\email{inhwa.song@princeton.edu}

\author{Kwangyoung Lee}
\orcid{0009-0006-4820-8486}
\affiliation{%
  \department{Department of Industrial Design}
  \institution{KAIST}
  \city{Daejeon}
  \country{Republic of Korea}
}
\email{kwangyoung@kaist.ac.kr}

\author{Janghee Cho}
\orcid{0000-0002-3193-2180}
\affiliation{%
  \department{Division of Industrial Design}
  \institution{National University of Singapore}
  \city{Singapore}
  \country{Singapore}
}
\email{jcho@nus.edu.sg}

\author{Amon Rapp}
\orcid{0000-0003-3855-9961}
\affiliation{%
  \department{Computer Science Department}
  \institution{University of Turin}
  \city{Torino}
  \country{Italy}
}
\email{amon.rapp@unito.it}

\author{Hwajung Hong}
\orcid{0000-0001-5268-3331}
\affiliation{%
  \department{Department of Industrial Design}
  \institution{KAIST}
  \city{Daejeon}
  \country{Republic of Korea}
}
\email{hwajung@kaist.ac.kr}


\begin{abstract}
    While Personal Informatics (PI) systems support behavior change, everyday well-being involves more than achieving individual target behaviors. It is shaped by cultural narratives that give actions meaning. In South Korea, the God-Saeng (God生) phenomenon—encompassing disciplined, collective, and publicly documented self-improvement practices—offers a lens into how well-being is negotiated in daily life. We conducted a 10-day probe (N=24) with bite-sized missions to examine how young adults engaged in God-Saeng. Participants relied on planning practices, accountability infrastructures, and datafication to stabilize themselves, yet these same routines also intensified pressures toward self-monitoring and performance. They navigated tensions between consistency and flexibility, authenticity and visibility, and productivity and broader values such as relationships, and reinterpreted ordinary activities through sociocultural contexts.
These insights suggest design opportunities for PI systems that move beyond tracking, toward digital instruments that help users negotiate tensions, make meaning, and reflexively understand how technologies participate in their culturally and existentially situated well-being.

\end{abstract}


\begin{teaserfigure}
\centering
    \includegraphics[width=\textwidth]{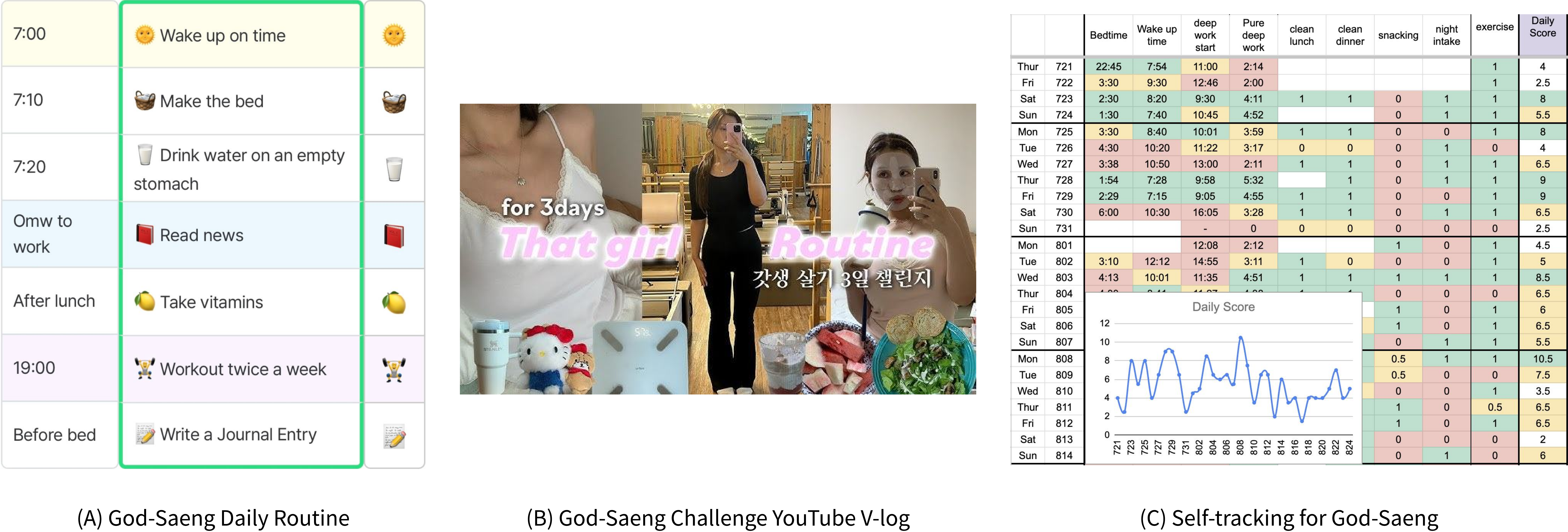}
    \caption[Composite of three visual examples of God-Saeng practices.
(A) Screenshot from the MyRoutine habit-tracking app showing a structured daily routine with activities and check-off icons.
(B) Still image from a YouTube vlog by “hyewonWee” showing a God-Saeng challenge day that includes a morning weigh-in, Pilates session, hydration with a tumbler, and evening gel mask care.
(C) Photo of a self-tracking log from an anonymous user displaying daily well-being scores captured through quantified metrics such as mood and activity levels.]{(A) Screenshot from the MyRoutine habit-tracking app~\cite{myroutine2020} illustrating a God-Saeng daily routine, (B) Video teaser of YouTuber ``hyewonWee'' (used with permission) showing a God-Saeng / That Girl daily challenge vlog—morning weigh-in, Pilates, hydration with a tumbler, and evening gel mask care, (C) Self-tracking practice shared by an anonymous user (used with permission)—daily well-being scores recorded through quantified metrics. }
    \label{fig:teaser}
\end{teaserfigure}


\maketitle
\section{Introduction}
Well-being is often treated as a universal goal, yet what it means to ``live well'' is always shaped by the cultural values and narratives of a society~\cite{compton2001values}. Everyday well-being is not simply the sum of discrete behaviors, but a process constructed through daily routines, social expectations, and personal values. \revised{These routines, which form the rhythm of everyday life, play a major role especially in orienting individuals amidst the growing uncertainty of the contemporary world, allowing them to regain a sense of steadiness and to interpret their own lives~\cite{van2009making, phipps2017routines}.}
Recent global trends such as social media challenges (e.g., \textit{That girl challenge}~\cite{stylist_thatgirl}, \textit{Study with me challenge}~\cite{lee2021personalizing}) illustrate how ideals of productivity, organization, and visible self-improvement have become dominant ways of expressing well-being in everyday life.

Within this broader trend, young adults in South Korea have localized these ideals through the notion of \textit{God-Saeng} (God生), which roughly translates as ``living a godly life.'' While often associated with disciplined routines such as waking up early, exercising, or studying, God-Saeng encompasses more than personal habit management. It involves active participation in challenges (e.g., daily journaling, study certification, or miracle morning routines), public documentation through social media, and a drive for continuous self-development. God-Saeng has been embraced as a way to pursue growth and articulate one’s identity in everyday life, yet it has also been critiqued as a form of performative well-being and even self-exploitation, reflecting how young adults negotiate what it means to live well. 

In HCI research, digital technologies that support behavior change are often studied under the frame of Personal Informatics (PI) or persuasive technologies~\cite{kersten2017personal,fritz2014persuasive,li2011personal}. 
While these systems have been successful in helping people track and achieve behavioral goals, they have paid less attention to how everyday well-being is constituted through cultural narratives, social expectations, and lived practices ~\cite{rapp2017know,rapp2023exploring,cho2022reflection,ayobi2016reflections}. As the God-Saeng phenomenon illustrates, practices of living well are not reducible to discrete behaviors or numerical targets, but are embedded in broader processes of growth, identity work, and even societal critique. This focus on numbers limits how PI systems can address the subjective, value-laden, and socially and culturally mediated dimensions of everyday well-being.

In this paper, we use the God-Saeng phenomenon as a lens to study how contemporary ideas of well-being are practiced, negotiated, and given meaning in everyday life by young adults. Thus, we ask:

\begin{itemize}
    \item How do young adults engage with God-Saeng practices in their everyday lives, and what cultural narratives shape their constructions of well-being?
    \item What tensions and negotiations emerge as individuals pursue growth and self-discipline through God-Saeng, particularly in relation to digital platforms?
    \item How can insights from God-Saeng inform the design of Personal Informatics systems that support meaning-making and value negotiation in everyday well-being?
\end{itemize}
\revised{To explore these questions, we conducted a multi-stage qualitative study with 24 young adults in South Korea. The study combined a group workshop, a 10-day in-situ probe, and a follow-up interview to access both the cultural narratives surrounding God-Saeng and the situated practices through which participants enacted and interpreted it. Central to this approach was a cultural-probe–inspired \textit{``bite-sized mission''} activity: participants first identified personally meaningful values during the workshop, then enacted small, self-authored missions grounded in those values over a 10-day period. This design allowed us to observe how participants navigated God-Saeng not only in discourse but through everyday actions, negotiations, and moments of tension embedded in their lived routines.}

\revised{Our findings show that while participants shared an aspirational image of God-Saeng centered on purposefulness, consistency, and self-improvement, its enactment relied heavily on sociotechnical infrastructures such as planning tools, accountability mechanisms, and datafied visibility. These systems helped participants stabilize themselves amid uncertainty, yet also intensified pressures toward self-monitoring and performance. Participants continually negotiated tensions between consistency and flexibility, authenticity and visibility, and productivity and broader values such as rest and relationships. Beyond these dynamics, participants also used the probe to reinterpret what it means to live well, grounding everyday moments in personally meaningful values.} 

Our study makes two contributions. First, we provide a \revised{situated} account of how \revised{young adults in South Korea mobilize sociotechnical routines—planning practices, accountability infrastructures, and datafied visibility—to navigate everyday precarity and construct personal well-being.} 
Second, we contribute design implications for PI systems, highlighting how they can move beyond tracking to support situated meaning-making and the ongoing negotiation of values within the everyday cultural contexts of users’ lives.
\section{Background and Related Work}
In this section, we cover the background on the (1) God-Saeng phenomenon, and the related work in the areas of \revised{(2) studying cultural practices in HCI,} (3) Personal Informatics for well-being and (4) contemporary design of Personal Informatics technologies.
\subsection{The God-Saeng Phenomenon}
\label{tab:background}

God-Saeng is a concept that emerged around the early 2020s in South Korea~\cite{GoogleTrendsGodSaeng}. The term combines `God', meaning god-like, with the Chinese character, `Saeng (生)', meaning life, which translates to `God-like life'~\cite{Song2024Godsaeng,Kwak2024Godsaeng}. God-Saeng refers to a lifestyle where individuals push themselves beyond their limits, constantly striving for productivity and self-discipline in diverse aspects of life. This notion involves maximizing every moment of the day—whether through waking up early, meticulously managing time and personal habits, or making time for self-care even in the midst of packed schedule.

The spread of God-Saeng lifestyle was driven by social media platforms like TikTok, Instagram, and YouTube, where individuals began to share their daily routines, often under the hashtag, \hashtag{God-Saeng Challenge}. 
These challenges involved publically and repetitively documenting daily activities such as going to the gym, waking up early, taking vitamin, and engaging in meditation. Key terms like \hashtag{God-Saeng Routine} (a set of habits or routines that someone follows for God-Saeng lifestyle) and \hashtag{God-Saenger} (someone who diligently lives by the values of God-Saeng) have emerged as a part of this phenomenon. 

\revised{In this study, we approach God-Saeng as a sociocultural trend through which young adults' everyday well-being practices are situated, interpreted, and made meaningful. This perspective enables us to explore how well-being is enacted through everyday practices that are shaped by cultural expectations and the technological infrastructures that mediate them.}
\begin{revisedenv}
\subsection{Studying Cultural Practices in HCI}
Researchers in HCI have long emphasized the importance of examining how technologies are embedded within the cultural, social, and material contexts of everyday life. Early critiques of universalist design assumptions argued that technologies cannot be understood, or made meaningful, outside the particular worlds in which they are taken up and interpreted~\cite{dourish2006implications}. This recognition has led to a rich body of work examining the cultural shaping of practices, values, and identities across diverse communities and socio-technical arrangements~\cite{pendse2023marginalization, pendse2022treatment, pendse2021can}.

An important theme in this tradition is the value of attending to what Tanenbaum and Tanenbaum~\cite{tanenbaum2018steampunk} describe as peripheral practices: \textit{``niche, unusual, marginalized, and/or highly specialized communities of practice whose study results in implications for HCI outside that community.''} Studying such cases serves not merely to document cultural specificity, but to provide conceptual and methodological resources for rethinking established assumptions about technology use. Peripheral practices promote defamiliarization, by making visible the implicit norms embedded in more dominant sociotechnical imaginaries and offer alternative models of meaning-making, motivation, and everyday life.

This cultural perspective parallels a growing interest in existential and phenomenologically informed HCI, which focuses on the exploration of phenomena within the individual’s universe of sense, foregrounding how technologies scaffold people’s attempts to construct coherence, agency, and continuity in the face of uncertainty~\cite{light2017design, rapp2023exploring, cho2024reinforcing}. Such work highlights that technology contributes to build sense around complex and precarious conditions, which are often tied to important personal, existential issues, like the need to be authentic, or to exert control over one's life ~\cite{kaptelinin, rapp2022}.

Taken together, these lines of work position cultural practices as meaningful sites for understanding how technologies intertwine with people's lived experience and support everyday meaning-making. Our study builds on this orientation by examining God-Saeng, a situated lifestyle discourse emerging among young adults in South Korea. Rather than treating it as a universal model of well-being, we approach God-Saeng as a culturally mediated practice that reveals how people organize routines and values, and make meaning around technology and important existential matters that shape their daily lives.

\end{revisedenv}

\subsection{Personal Informatics for Well-being}
Personal Informatics (PI) systems are tools designed for individuals to collect, analyze, and reflect on personal data, aiming for well-being~\cite{li2010stage}. Advancements in PI systems recently have expanded into diverse domains such as mental health, diet, sleep, and personal finances~\cite{epstein2020mapping}. This growth reflects a broader shift in PI technologies, increasingly recognized not just as tools for data collection, but as catalysts for behavior change, driven by self-improvement and deeper self-awareness~\cite{choe2014understanding}. Over time, the conceptualization and application of PI systems have evolved. Early models, such as the stage-based model by Li et al.~\cite{li2010stage}, emphasized a linear progression through stages of data preparation, collection, integration, reflection, and action, primarily focusing on the interaction between a user and the PI system for behavior change. Subsequent frameworks, like Epstein et al.'s Lived Informatics Model~\cite{epstein2015lived}, Niess \& Woźniak’s Tracker Goal Evolution Model~\cite{niess2018supporting}, and Lu et al.'s Socially Sustained Self-Tracking Model~\cite{lu2021model} expanded this perspective by integrating everyday life experiences, acknowledging the diverse contexts and personal circumstances in which PI tools are used. 

Recently, Rapp et al.~\cite{rapp2023exploring} introduced the integration of an existential approach to behavior change, emphasizing holistic understanding and sense-making. This model advocates for technology designs that support users in making sense of their behaviors in the context of broader existential matters. Current technologies often fall short in facilitating this deeper level of reflection and insight generation based on lived experiences~\cite{cho2022reflection}. 
The existential model highlights the importance of connecting behavior change to personal meanings and life contexts~\cite{rapp2023exploring}. 
Building on this trajectory, our study examines how a culturally specific well-being narrative—God-Saeng—shapes everyday practices and personal sense-making, extending PI research beyond individual behavior change.

\subsection{Contemporary Design of Personal Informatics Technologies}

PI systems help individuals collect personal data, gain insights, and achieve behavioral goals~\cite{choe2014understanding} through diverse design features, such as providing statistics and visualizing progress~\cite{cuttone2013mobile, cuttone2014four}, gamification (e.g., providing rewards)~\cite{rapp2018gamification}, and offering practical tips for setting tangible well-being goals~\cite{ekhtiar2023goals}. These features have demonstrated a significant contribution to improving well-being across diverse populations and contexts~\cite{kim2019toward, harrington2018informing}.   

However, in recent years, there has been growing discussion in the HCI and CSCW fields around the limitations of current PI technology designs~\cite{cho2022reflection, rapp2017know, rapp2023exploring, rapp2023wearable}.
Many of these systems focus primarily on modifying behavior, often exclusively prioritizing measurable outcomes (e.g., number of steps taken, calories burned), while overlooking the deeply subjective and meaning-laden experiences of behavior change~\cite{rapp2023wearable}. As a result, the focus on metrics can even reduce the enjoyment of those activities~\cite{etkin2016hidden}.
Furthermore, quantification of well-being can sometimes lead to reinforcement of societal ideals~\cite{sointu2005rise, white2017relational}, in ways that do not fully reflect the complex nature of well-being~\cite{cho2022reflection}.

In our research, we aim to acknowledge not only the quantifiable aspects of well-being but also its intangible dimensions, such as the importance of personal values and everyday contexts, which contribute to a deeper discourse around well-being.

\section{Method}
\begin{table}[t]
\sffamily
\small
\def\arraystretch{1.1}\setlength{\tabcolsep}{0.5em}
\caption[Table listing 24 participants organized in nine groups (G1–G9). Each row shows participant ID, age (20–26), gender, and three self-chosen value keywords from the workshop. Values span personal growth, relationships, family, nature, relaxation, achievement, and similar themes.]{Demographic information of the participants and the value keywords elicited during the workshop activity.}~\label{tab:demographic}

\begin{tabular}{|
>{\centering\arraybackslash}m{0.1\columnwidth}!{\color{tablegrayline}\vrule}c!{\color{tablegrayline}\vrule}c!{\color{tablegrayline}\vrule}m{0.1\columnwidth}!{\color{tablegrayline}\vrule}m{0.4\columnwidth}|}
\hline
\rowcolor{tableheader}
\makecell[c]{\textbf{Group}} & \textbf{Participant} & \textbf{Age} & \textbf{Gender} &\textbf{Value Keywords} \\ \hline
\multirow{3}{*}{G1} & P1 & 21  & Female & Growth, Relaxation,\newline{}Gratitude \\ 
 & P2          & 21  & Female & Happiness, Achievement, Challenge \\
 & P3          & 20  & Female& Achievement, Relaxation,\newline{} Relationships\\ \hline
\multirow{3}{*}{G2} & P4          & 23  & Male & People, Daily Life, Events\\
 & P5          & 20  & Female& Friends, Family, Scenery \\
 & P6          & 20  & Female& Healing, Reflection, Love\\ \hline
\multirow{3}{*}{G3} & P7          & 20  & Female& Small Joys, Achievement, Meaningful Happiness\\
 & P8          & 21  & Female& Family, Achievement, Entertainment \\
 & P9          & 20  & Female& Stability (Healing), Achievement, Saving\\ \hline
\multirow{2}{*}{G4} & P10         & 26  & Male& Nature, Happiness, Likes\\
 & P11         & 25  & Female& Nature, Excitement, Saving\\ \hline
\multirow{2}{*}{G5} & P12         & 21  & Female& Relationships, Sense of Achievement, Emotion\\
 & P13         & 23  & Male& Relaxation, Living Life Earnestly, Relationships\\ \hline
\multirow{4}{*}{G6} & P14         & 20  & Male& Relationships (Love), Aesthetics, Achievement\\
 & P15          & 22  & Female& Research, Reflection, People\\
 & P16          & 21  & Female& Relaxation, Happiness, Emotions\\
 & P17         & 21  & Male& Self, Relationships, Achievement\\ \hline
\multirow{2}{*}{G7} & P18         & 24  & Male& Relationships, Achievement, Relaxation\\
 & P19         & 24  & Female& Fun Things, Experiences, Warmth\\ \hline
\multirow{2}{*}{G8} & P20         & 20  & Male& Relationships, Courage, Enjoyment\\
 & P21         & 24  & Male & Family, Happiness, Effort\\ \hline
\multirow{3}{*}{G9} & P22         & 22  & Male& People, Relationships, Stability\\
 & P23         & 24  & Male& Family, Self, Art\\
 & P24         & 23  & Female& Relaxation, Achievement, Relationships\\ \hline
\end{tabular}
\end{table}

\subsection{Participants and Recruitment}

To explore how God-Saeng is experienced, interpreted, and potentially contested in the everyday meaning-making of young adults, we recruited participants from \textit{Generation Z \revised{(age 20-26)} in South Korea}, a demographic that most actively engages with and circulates this cultural trend~\cite{song2023global, lee2024owoonwan}. 
\revised{Because our goal was to understand how God-Saeng is lived and attempted---not only how it is successfully achieved---we focused on \textit{individuals who aspired to practice God-Saeng, while intentionally sampling diverse practice experience and styles of engagement.}}

We employed a multi-channel recruitment strategy, including flyers, social media posts, word-of-mouth, a local university’s online forum, and active office worker communities. 
The screening survey collected basic demographics (age, gender\revised{, occupation (if applicable)) and participants' God-Saeng engagement profiles, composed of (1) their engagement trajectory prompted through an open-ended description accompanied by examples (\textit{e.g.,} ``I consistently practice God-Saeng routines,'' ``I practice them intermittently,'' ``I used to practice them but not recently,'' ``I want to practice God-Saeng but struggle to maintain it'')} to illustrate the range of possible relationships to God-Saeng, and (2) the practices they have tried when attempting to live a God-Saeng lifestyle, described through an open-ended question about routines, challenges, apps, or strategies they have engaged in. \finalrevised{We opted for open-ended descriptions to capture participants’ first-person interpretations of their engagement with the phenomenon, in line with the phenomenological and existential perspectives we adopted ~\cite{rapp2022, rapp2023exploring}, which treat engagement as a lived, situated phenomenon rather than a fixed, externally measurable variable. Engagement was thus operationalized through participants’ self-described trajectories and practices, using guided prompts to ensure interpretive coherence across responses.} \revised{These questions enabled us to purposefully sample participants across an engagement spectrum ~\cite{marshall1996sampling}.}

From 131 screening responses, we selected 24 participants \deleted{(ages 20–26; 8 male, 16 female), all Korean nationals,} varying in gender, occupation (e.g., students, workers, and those in transition), and God-Saeng engagement profiles.
We invited participants to register in pre-existing friendship groups of two to four, 
\revised{allowing group size to vary naturally based on existing relationships: we expected that familiar peers would offer a comfortable context for discussing meaningful experiences and sharing interpretations. This choice is grounded in prior work in qualitative and HCI research, which notes that when studying culturally shaped everyday practices, familiar peers can provide a supportive context that enables richer recall and articulation of shared norms, colloquial expressions, and lived experience~\cite{morgan1997focus}.}

Participants were compensated 50,000 KRW (approx. \$40 USD) for their participation across all phases of the study: a one-day group workshop, a 10-day individual task, and a follow-up interview.
The study received Institutional Review Board (IRB) approval prior to recruitment, and informed consent was obtained from all participants. We emphasized voluntary participation, and participants could withdraw at any time. We clearly communicated that they were not expected to meet any fixed performance standards during the study.

\subsection{Probe Design}
\revised{The study design was guided by the goal of understanding how young adults \textit{situate} their everyday well-being practices within the broader cultural narratives of God-Saeng. We used a cultural probe–inspired design to access the tacit and situated aspects of participants’ everyday lived experience. This approach allowed us to explore not only what participants say about God-Saeng, but also how they enact or experience it in their daily lives. 
We structured the study in three parts: (1) a group workshop, (2) 10-day period of individually enacted missions (which we later phrase as bite-sized missions), and (3) a follow-up interview.} 
\begin{figure*}[t]
\centering
\includegraphics[width=\textwidth]{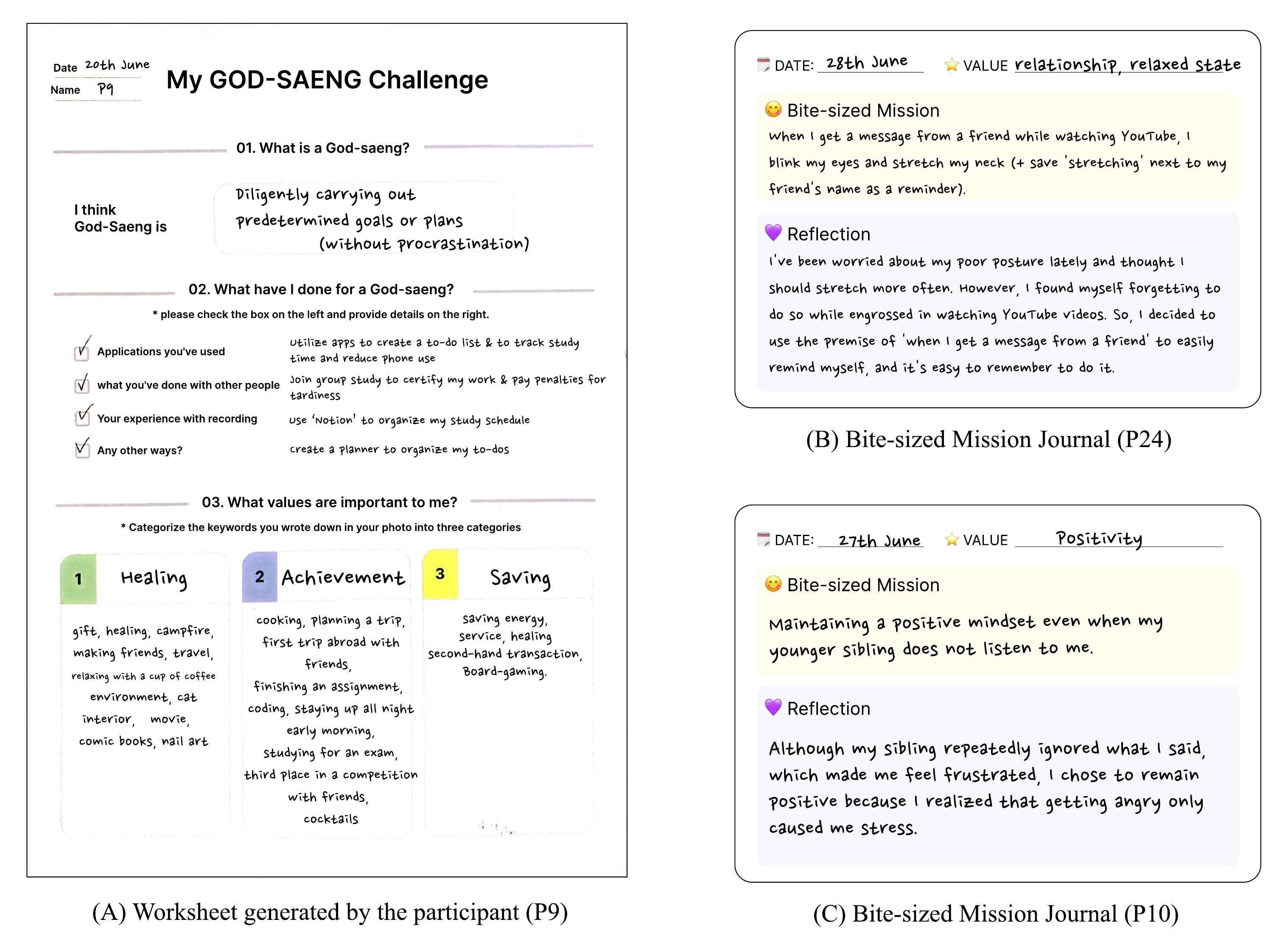}
\caption[Three-part composite image showing participant-generated materials.
(A) A handwritten workshop worksheet where participant P9 defines “God-Saeng,” lists personal approaches, and identifies important values with keywords.
(B) A bite-sized mission journal entry from participant P24 describing a mission to maintain a positive mindset when their younger sibling ignored them, including a short reflective note.
(C) A bite-sized mission journal entry from participant P10 outlining a stretching reminder strategy triggered by receiving a friend’s message while watching YouTube, accompanied by a reflection on posture habits.]{(A) Translated version of participant-generated material. Participants filled out a worksheet about their own definition of God-Saeng, practices. (B, C) They created ‘bite-sized missions’ grounded in the values they have identified during the workshop.}
\label{fig:material}
\end{figure*}
\subsubsection{Group Workshop}
\revised{We began the study with a 60-minute in-person group workshop designed to understand participants’ shared and individual interpretations of God-Saeng, as well as the everyday practices through which they have attempted to enact it.}

\revised{The workshop began with activities prompting participants to define God-Saeng in their own terms. Using a worksheet (Fig. ~\ref{fig:material}A), each participant wrote a short description of what they believed constitutes a ``God-Saeng'' life: here we wanted to capture participants' definition of God-Saeng before discussion where they could influence each other. Then participants followed by an open-ended group conversation where they shared and elaborated the definitions they have written, and discussed around how God-Saeng manifests in social media, peer culture, etc.}

\revised{Next, participants described the concrete practices they had previously attempted in pursuit of God-Saeng. Using the worksheet (Fig.~\ref{fig:material}A) that prompted them to reflect on categories such as ``application used,'' ``social practices,'' ``documentation practices,'' and ``other practices,'' each participant first wrote down the strategies they had tried. Again, at this stage we aimed to capture their individual idiosyncratic experiences. Participants then shared these experiences with the group, elaborating on motivations, challenges, and the circumstances of their attempts. Through the discussion, we aimed to help participants deepen their understanding of the phenomenon and their own attempts, by identifying similarities and differences with the experiences of others. This activity surfaced the diverse ways God-Saeng is enacted and how it intersects with participants' lived constraints.}

\revised{The latter part of the workshop introduced participants to the upcoming phase of the study: the 10-day bite-sized missions. As part of this transition, during the workshop, participants individually completed a value-keyword identification activity designed to ground the 10-day missions in personally meaningful themes. The details of this activity is described in section~\ref{sec:bsm}} 
\subsubsection{Bite-sized Mission Phase}
\label{sec:bsm}
The second phase of the study invited participants to engage in a 10-day series of bite-sized missions---small, self-authored actions grounded in the values they had identified during the group workshop. 

\begin{revisedenv}

\ipstart{Identifying Value Keywords}
To prepare participants for the mission phase, the workshop included a structured individual activity designed to help them reflect on the values that meaningfully shaped their everyday experiences. 
To operationalize this process, we incorporated participant-generated photographs as reflective prompts, following research in HCI and cultural studies that show that images can evoke tacit memories and emotions that verbal recall alone often fails to surface~\cite{liebenberg2012analysing, mitchell2011doing}.

Five days before the workshop, participants were asked to collect and submit 10-20 photographs that captured moments they personally considered meaningful in their daily lives. These could include people, objects, places, ambient scenes, or everyday situations—whether newly taken or drawn from their existing camera rolls. The purpose of the photographs was not to analyze their visual content per se, but to serve as memory prompts that could evoke the situated experience.

At the workshop, the research team provided printed versions of each participant’s submitted photos, leaving space below for annotations. Participants first labeled each photo with the rich elements that made the moment meaningful, such as relational dynamics, sensory qualities, environmental features, feelings, or other contextual aspects. They then clustered these annotated photos into three thematic groups and assigned each group a value keyword that best represented its core elements. (See the lower part of Fig. ~\ref{fig:material}A). 

Through this process, each participant defined three value keywords that represented themes consistently present in moments they found meaningful (e.g., connection, growth, productivity, calm). Immediately following this activity, we introduced the bite-sized missions and guided participants to use their own value keywords in constructing their 10-day missions. 

\ipstart{Bite-sized Missions} 
The bite-sized mission phase was designed to access aspects of everyday meaning-making that are often difficult to surface through interviews alone, allowing us to observe how participants' interpretations of God-Saeng took shape through situated action.

Here, the value keywords defined in the previous activity served as interpretive anchors that participants could flexibly draw upon when crafting their missions, rather than as categorical labels, through which they could frame and explore moments of their everyday lives. By grounding each mission in a value that participants identified as personally meaningful, the activity encouraged them to examine not only what they were doing, but \textit{why} it mattered to them and how these moments interacted with broader cultural narratives of God-Saeng.

Participants were thus invited to self-define the missions. Each mission was recorded on a printed card consisting of three fields: (1) value keyword(s) selected by the participant, (2) a short description of an enactment inspired by that value, and (3) an open reflection box where participants documented thoughts, emotions, or contextual details after doing the mission (See Fig. ~\ref{fig:material} B, C). 

Participants could complete one to three missions per day.
We did not define what a mission should look like, and there were no expectations regarding difficulty, productivity, or performance. In this spirit, we noted that missions could take many forms, for example, ranging from familiar routines to new attempts, from concrete tasks to small shifts in awareness, and from brief, mundane acts to more experimental or reflective moments.

\ipstart{10-day Mission Execution}
After defining the missions, participants carried out them independently for ten consecutive days. 
Participants could complete the missions at any point during the day, and were asked only to send a photo of their completed card(s) to the research team via a private messaging channel by the end of the day.
\end{revisedenv}
\subsubsection{Follow-Up Interview}
After completing the 10-day challenge, participants reconvened in their original peer groups for a 60-minute group interview. These interviews were designed to facilitate reflection on participants’ experiences with the bite-sized missions and to explore how their everyday meaning-making related to or diverged from the God-Saeng narrative.
\revised{Building on the shared foundation established in the workshop,} participants were invited to revisit specific missions they had completed, share memorable moments that felt particularly meaningful or challenging, and \revised{articulate how their reflections evolved over the 10 days. We also prompted them to consider how these everyday experiences connected back to their earlier conversations about God-Saeng.} 
\subsection{Data Analysis}
\begin{figure}[h]
\centering
\includegraphics[width=\columnwidth]{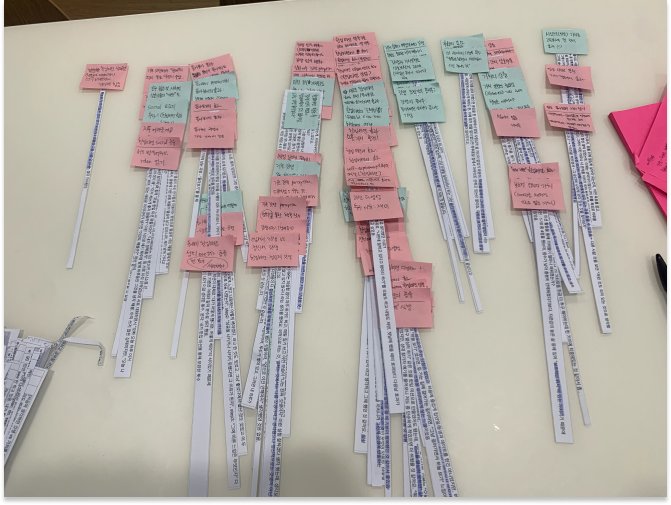}
\caption{Photograph of the manual coding process, where themes and categories were developed by analyzing participant responses from the workshop discussions, journal entries, and interviews. Post-it notes represent the coded data, grouped into related themes for further analysis and refinement.}
\label{fig:analysis}
\end{figure}
We analyzed data from three sources: (1) workshop transcripts and worksheets, where participants articulated their understandings of God-Saeng and surfaced personal values, (2) 10-day bite-sized mission cards, which documented the missions they designed, enacted, and reflected upon, and (3) follow-up group interviews, where participants reflected on their practices and related to broader cultural narratives. 

\revised{All group workshops and follow-up interviews were recorded and fully transcribed in Korean. Workshop sheets and bite-sized mission cards (participants' written descriptions of their value keywords, mission intentions, and daily reflections) were digitized in textual data. Materials from the value-keyword activity (e.g., labeled photo annotations) were used as contextual artifacts that informed code development.}

\revised{We approached the data through an interpretivist lens, focusing on how participants made sense of their routines, struggles, and aspirations within a broader sociotechnical and cultural context~\cite{marshall1996sampling}. Our thematic analysis was phenomenologically oriented, attending to how meanings were constructed through lived experience rather than inferred from discrete behaviors~\cite{sundler2019qualitative}.}

\revised{The open coding was done iteratively. In the first cycle, three researchers conducted line-by-line open coding across workshop transcripts, mission reflections, and interview data. In the second cycle, codes were clustered into preliminary categories such as how participants structured their routines, how they interacted with technological supports, and how they reflected on the meanings of their everyday actions. Then, we analyzed the workshop discussions, mission cards, and follow-up interviews in relation to one another, examining how participants’ interpretations and practices evolved across the study. Throughout analysis, researchers met regularly to review analytic memos, challenge assumptions, and refine themes. As is common in qualitative research adopting an interpretive approach ~\cite{Yardley01032000, braun2013successful, harry}, no numerical reliability rating is reported, because our aim was to reach an intersubjective consensus, where each point of difference was debated until the researchers agreed on appropriate usage of the identified codes ~\cite{harry}. Discrepancies were then resolved through discussion grounded in the data. Similarly, we chose not to report the exact number of participants expressing a given perspective in the findings, as we did not want to suggest that these findings could be evaluated within a scientific frame that treats numbers as key indicators of validity and generalizability. Such a frame belongs more to the positivistic tradition than to the interpretivist one, which we embraced in this research~\cite{weiss1995learning, maxwellqual, patton}.}
\section{Findings}
In this section, we present how participants 1) engaged with everyday practices, \revised{sociotechnical infrastructure, and} cultural narratives of the God-Saeng phenomenon (RQ1), 2) navigated the tensions and negotiations that emerged in pursuing God-Saeng (RQ2), and 3) reinterpreted and reshaped their well-being through values and personal narratives (RQ3).
 
\subsection{Perceptions and Practices of God-Saeng}

\subsubsection{Interpreting God-Saeng}
\begingroup
\begin{table*}[t]
\sffamily
\small
\def\arraystretch{1.2}
\setlength{\tabcolsep}{0.5em}
\centering
\caption[Interpretive themes derived from participants' descriptions of the God-Saeng concept. Each theme summarizes how participants articulated what it means to ``live a God-Saeng life.'']{\revised{Interpretive themes derived from participants’ descriptions of how they understand God-Saeng.}}
\label{tab:godsaeng_themes}

\begin{revisedenv}
\begin{tabular}{|m{0.15\textwidth}!{\color{tablegrayline}\vrule}p{0.3\textwidth}!{\color{tablegrayline}\vrule}p{0.4\textwidth}|} 
\hline
\rowcolor{tableheader}
\textbf{Theme} & \textbf{Description} & \textbf{Words of Participants} \\ \hline

Purposefulness \& \newline{} Performativity &
Living with direction and intention. Performing everyday time, resources, and actions with what was planned ahead. & 
\textit{``Living the day the way I intended it. (P14)''} \newline
\textit{``Having days that feel achieved today and motivated for tomorrow. (P16)''} \newline
\textit{``Finding happiness through a sense of achievement. (P5)''} \newline
\textit{``Living un-regrettable days. (P2, P13)''} \newline
\textit{``Managing my energy and time efficiently the way I intend. (P15)''} 
\\ \hline

Diligency \& \newline{}Consistency &
Sustaining disciplined routines—showing up daily, doing what needs to be done without delay, and maintaining commitment through persistence. & 
\textit{``Continued days of putting my plans into action without postponing them. (P7, P9)''} \newline
\textit{``Building healthy routines and sticking to them everyday (P6). ''}\newline
\textit{``Doing things every day, steadily. (P22)''} \newline
\textit{``Being diligent and persistent in what I’m responsible for. (P8)''} 
 \\ \hline

Self-improvement \& \newline{}Productivity &
Pursuing personal growth and becoming a better version of oneself through self-development, healthy habits, and a sense of progress. 
& 
\textit{``A life where I can make the most of myself (P12)''} \newline
\textit{``Not neglecting my own growth or self-development. (P8)''} \newline
\textit{``Filling my day with time that isn’t wasted. (P23)''} \newline
\textit{``Living diligently and passionately, feeling like I can do things well. (P19)''} \newline
\textit{``Becoming someone who takes care of myself. (P11)''} \newline
\textit{``Creating routines that help me become better. (P24)''} \\ \hline
\end{tabular}
\end{revisedenv}
\end{table*}
\endgroup

\begin{revisedenv}
Participants described a range of interpretations of God-Saeng, yet their explanations converged around a shared aspirational image. As summarized in Table \ref{tab:godsaeng_themes}, participants framed God-Saeng as a way of living that is purposeful, consistent, and oriented toward continuous self-improvement. Across the workshops and interviews, purposefulness emerged as an important theme, an orientation toward living with intention rather than drifting through daily life. P4 explained, \textit{``If I don’t make plans, it becomes hard to live with any sense of purpose—so I try to plan intentionally, even my personality isn't born to make plans.''} Consistency and diligence similarly defined what most participants regarded as ``living well.'' As P20 noted, \textit{``The opposite of God-Saeng might be just lying in bed watching YouTube all day.''} Finally, many participants associated God-Saeng with a strong desire for self-improvement, projecting an image of someone who excels well in every domain of life. P13 described an ideal God-Saeng image as, \textit{``someone like a career woman who is excellent in her work, has many friends, is good with relationships, and has a well-shaped body.''}

While this aspirational image was consistent, its enactment in everyday life sometimes revealed boundaries between what counted as God-Saeng and what did not. Participants often drew distinctions between values associated with God-Saeng—such as productivity, discipline, and goal pursuit—and other meaningful aspects of life that were more relational, leisurely, or restorative. P7 noted that pursuing God-Saeng during the semester came at a cost: \textit{``Trying to live a God-Saeng lifestyle really damaged my friendships. I only saw my university friends whom I was studying with, and even when my high school friends asked to meet, I never had the time.''} For some, domains like rest or sleep felt incompatible with the demands of God-Saeng. As P6 noted, \textit{``Once I make a plan, I feel like I have to follow it no matter what—even if that means not sleeping until everything is done. So sleep often feels contradictory to God-Saeng.''} In these accounts, God-Saeng ideals were holistic in theory yet narrowed in practice, often privileging achievement-related activities while sidelining other forms of well-being.

Beyond how participants understood God-Saeng, they also varied in how they positioned themselves in relation to this ideal. A few described that, while challenging, they were quite confident that they were disciplined God-Saengers. Some described a metaphor of engaging `cyclically,' adopting God-Saeng practices only during certain periods such as the \textit{``academic semester (P7).''} Several participants believed that consistent effort would eventually enable them to \textit{``live intentionally and keep growing (P24).''} P19 admitted to countless attempts at a God-Saeng lifestyle, joking, \textit{``I guess this is my 99th God-Saeng trial? Haha.''} In contrast, a number of participants described self-blame (P21) and chronic feelings (P16) of falling short despite their high aspiration for the God-Saeng lifestyle, interpreting lapses as \textit{``personal inadequacy rather than circumstantial difficulty (P21).''} Still others had been deeply invested in God-Saeng in the past but spoke with growing skepticism, questioning whether its demands were compatible with \textit{``sustainable well-being (P10).''} 

\todo{short paragraph about uncertainty related quotes + not a sustainable livable sense of self}

Together, these accounts show that while participants shared a broad consensus on what God-Saeng signifies, their lived relationships to this ideal were not always uniform. God-Saeng operated simultaneously as an aspirational image, a source of pressure, a moral orientation for daily life, and something participants continually negotiated. 

\end{revisedenv}
\begin{revisedenv}
\subsubsection{Enacting God-Saeng Through Socio-Technical Practices}
Participants enacted God-Saeng through everyday routines by assembling a variety of technological and social infrastructures to sustain their desired way of living. These infrastructures blended technological components (e.g., planning tools, checklist apps, timestamp-based verification systems) that structured daily routines with social and procedural components (e.g., peer check-ins, shared rules, challenge commitments) that activated and reinforced them, together stabilizing participants’ desired ways of living. Three patterns highlight how participants practiced God-Saeng in situated and practical ways.

\ipstart{Scripting the Day: Planning and Routine Construction}
To enact the purposeful life associated with God-Saeng, participants invested effort in structuring, breaking down, and operationalizing their plans in everyday life. Participants described building layered planning systems, such as \textit{``yearly visions, quarterly targets, weekly structures, and daily checklists (P12, P14)''}. Some drew parallels to project-management workflows they had encountered at work. For example, P16, who was working at a startup company explained, \textit{``I try to manage my life the same way my team does the project management. I found applying methods like sprint cycles and scrum helps me manage my plans and routines.''} P15 used Notion\footnote{Notion is an all-purpose workspace for writing, scheduling, and organizing personal or collaborative projects. https://www.notion.com} to visually modularize life areas (e.g., health, career, emotions) into nested goals and sub-tasks (e.g., maintaining a regular sleep cycle, focusing for three hours daily, keeping diary), noting how digital tools made it easier to \textit{``group, rearrange, and break things down''} by making them visible and more concrete.

In particular, participants emphasized that planning was not only about setting direction, but about making intentions actionable. For example, P5 described how she used TimeTree~\footnote{TimeTree is a shared calendar application that supports routine management through scheduling, color-coding, reminders, and collaborative planning features.
https://timetreeapp.com} to manage her monthly goals:
\begin{quote}
    \textit{``First I set my monthly goals and break them into big tasks. Then I put those tasks directly on the monthly calendar with specific dates. After that, I divide each big task into smaller to-dos and attach checklists to each scheduled item.''} (P5)
\end{quote}
Several others used similar combinations of calendar apps, checklists, and progress logs to continually refine, schedule, visualize, and monitor their plans, turning intention into a sequence of  steps that could orient them in their everyday life.

For some, planning also functioned as a form of self-imposed obligation. P8 described creating what she called \textit{``satisfaction debt''} by setting plans in advance so that she would \textit{``owe it to myself to follow through.''} These plans later became tied to various accountability and datafication mechanisms which we describe next.  

\ipstart{Technology as Accountability Infrastructure}
Many participants relied on technology not merely to track behaviors but to create a system of external accountability that would hold them to their intentions. 
These systems often combined digital platforms, social groups, and monetary stakes, transforming personal routines into commitment that were socially or technologically enforced.

More than half of the participants used social platforms such as group chats to build structured accountability groups. A Slack-based retrospection group was a notable example. 
G6 (P14-17) were participating in a nine-week retrospection Slack group with roughly 10 peers where members were required to submit a 400-word weekly reflection and post comments on others’ entries; failure triggered an automatic deduction from a pre-paid deposit. They described these infrastructures as \textit{``forcing me to keep going even when motivation drops (P12),''} noting that accountability was produced not only through peers but also through the system’s built-in rules, deadlines, and penalties. They also added ``buddy systems,'' pairing members to review each other’s reflections and ensure mutual adherence, effectively distributing the labor of discipline across social ties. 

Beyond weekly retrospection groups, there was a range of infrastructures that participants had constructed, centered around social technologies. Some participants formed group chats where members checked in with one another to maintain accountability. In these spaces, they either shared the same daily routines and monitored each other’s progress, or each person followed their own set of routines (e.g., a list of things they planned to do every day) and simply reported back to stay committed. Some referred to this format as a ``God-Saeng study.'' P12 explained that, depending on their needs at the time, they would recruit people to form such study groups. As she described: \textit{``I’ve recruited members for wake-up studies and daily gym check-in studies. Sometimes friends would join, and other times I would post on university online communities to find study partners.''} In contrast, P13 noted that they personally found it burdensome to set rules and maintain them, so they usually look for groups that others have already organized rather than creating their own. 

Apps designed explicitly for behavioral enforcement played similar roles for many participants. 
For P4, Alarmi app\footnote{Alarmy is a smartphone alarm application that enforces task-based dismissal (e.g., puzzles, photo tasks) to ensure wake-up compliance. https://alar.my} enforced wake-up routines through tasks like math puzzles and memory games that had to be solved to silence the alarm, creating a technological barrier against oversleeping. Even comparatively simple tools were appropriated as accountability systems. For more than a third of participants, Todomate app\footnote{Todomate is a checklist-based social task management app that helps users structure daily routines, track and share progress through recurring to-dos. https://play.google.com/store/apps/details?id=com.undefined.mate} was adopted to share daily checklists with friends and respond to each checkbox with emoji reactions. P7 described the sense that \textit{``if I don't check it off, my friends will see,''} turning a private to-do list into a shared commitment. Others leveraged broader audiences for accountability. For instance, P19 created a public blog dedicated to \textit{``declaring routine goals,''} using posts as a way to signal commitments to followers and hold herself accountable for completing routines. 

\ipstart{Performing Visibility and Self-Verification}
Beyond sustaining routines, participants’ God-Saeng practices were deeply intertwined with the need to verify, to oneself and/or to others, that one was living according to their intended routines and plans. 
Technologies played a central role in this process, providing material traces that rendered everyday actions verifiable.
Across the study, participants assembled diverse verification mechanisms that transformed their routines into observable accomplishments.

One set of practices involved \textit{app-based simple materialization of progress}, where everyday routines were translated into streak counts or monthly visualizations. For some, the specific routine mattered less than the record it produced. As P12 who used MyRoutine\footnote{MyRoutine is a habit-tracking mobile app that allows users to build recurring routines, receive reminders, and monitor streaks over time. https://myroutine.today/en} explained, \textit{``Looking back, I think consistently keeping up with the routine itself becomes my purpose. The sustained streak in the MyRoutine app helps me feel that I am in control of myself,''} showing how the representational artifact, not merely the action, became meaningful.

Another dominant practice across participants relied on \textit{timestamp-based proof-of-doing}. In wake-up study groups (P2, P4, P9, P12), routine-sharing group chats (P7, P12, P15-17, P19), and app-based challenges like Challengers~\footnote{Challengers is a habit-building app where users wager money and verify completion through timestamped proof. https://play.google.com/store/apps/details?id=kr.co.whitecube.chlngers} (P3, P12, P14), participants used timestamp apps to certify punctuality or completion. These systems provided ``tamper-proof'' evidence that an action happened at a specific moment. P14 described how his engagement was shaped across varying Challengers challenges: 

\begin{quote}
\textit{``For the `8 a.m. wake-up' challenge, I had to go to the bathroom, turn on the faucet, and take a TimeStamp photo with your hand under the water. For the ‘read five days a week’ challenge, the rule was to verify by opening the book and taking a photo. Having my hand under the faucet really made me get up, but on busy days, I sometimes just opened the book and took a picture without reading.''}
\end{quote}

Participants also engaged in \textit{real-time forms of visibility}. Apps such as Yeolpumta\footnote{Yeolpumta is an app that measures study time, blocks distracting apps, and displays live study durations to peers in a virtual online space. https://www.yeolpumta.com}, widely used by a third of participants, displayed users’ continuous study time within shared online rooms. 
Some participants similarly used Zoom study sessions with their friends with cameras left on to maintain co-presence and a light social pressure. Here, \textit{``being shown on the camera (P13)''} during the study session served as a form of social verification of one's effort. 
At the same time, participants were acutely aware of the system’s boundaries. P9 mentioned doubting a friend’s \textit{``10-hour streak in Yeolpumta,''} joking that they must have \textit{``found a loophole.''} 
Some participants used \textit{non-digital and embodied} forms of visibility. Participants described joining running crews (P14) or attending in-person morning gatherings (P12, P14), where physical presence functioned as a form of self-verification. 

Across these participants' examples, verification was not merely a by-product of action but became an opportunity for meaning-making and for finding a sense of control over oneself, as P12 noted, \textit{``Seeing and reminding the accomplishments of the daily routines that I have established are some of the only sources of satisfaction I can find these days. To be honest, there aren’t many places where I can get that sense of accomplishment. There's not much I have in control. I have to create these kind of opportunities of satisfaction by myself.''} 

Yet the same infrastructures that stabilized participants’ routines also carried emotional risks. When participants struggled to keep up with accountability systems, the lack of verification or missed check-ins was often interpreted as a personal shortcoming. After repeatedly falling behind in a God-Saeng group chat, P20 shared, \textit{``Lying on the bed when I was supposed to wake up and check-in to the group chat, I felt like an immoral person.''} Others pointed out the rigidity of these systems. P7 explained that while daily routines offered stability during a stressful job search, they also produced burden: \textit{``After graduating college, during my job-seeking phase, everything seemed uncertain. Using tracking tools to manage my daily routines helps me feel a small sense of achievement every day. At the same time, I have to admit that I often push myself too hard just to stay on track, which I realize is ironic. I'm not sure what to do.''} P13 also reflected, \textit{``Using a commitment device for my daily goal of walking 10,000 steps, where failing even by a minute results in a fine, forces me to adhere strictly to these criteria. It makes the process very rigid.''} These sociotechnical arrangements thus acted as non-neutral infrastructures---supporting consistency and self-recognition for some, while amplifying anxiety or self-blame for others. 
\end{revisedenv}
\subsection{Negotiating Tensions in God-Saeng}
\subsubsection{Consistency vs. Flexibility}
The journey toward God-Saeng was often described as a cycle of persistence, failure, and renewal, where consistency itself became a mark of discipline and achievement.
P9 explained, \textit{``For it to feel like real God-Saeng, I have to complete at least one small mission every day. A single burst of effort isn’t enough—it’s the daily commitment that matters.''}
Similarly, P7 said, \textit{``I tried to do at least one thing I planned each day and felt proud when I kept that promise to myself.''}
P16 echoed this view, treating the recurring pattern of \textit{``fail and try again''} as an achievement in itself.

At the same time, strict consistency often collided with everyday contingencies, prompting flexible adaptations and self-forgiveness.
P11 resisted the pressure to follow rigid routines, advocating for a more subjective approach to happiness and well-being, divorced from societal dictates. 
Probe reflections reinforced the adaptive mindset. 
For instance, P1 set a 3 km run for the bite-sized mission, but alternated between running and walking in the heat, noting \textit{``it was hard, but I enjoyed it anyway.''} 
P3 similarly said, \textit{``I couldn’t complete everything for reasons beyond my control, but I counted housework as my exercise and felt okay about it.''} 
Others found that journaling revealed unexpected insights. P13 mentioned, \textit{``When plans failed I used to feel depressed, but while writing reflections I realized good days aren’t only when things go as planned.''}

Flexibility also helped preserve meaning when routines risked becoming rote.
P12 observed, \textit{``The moment I felt I had to continue the same mission every day to fulfill a value, it started feeling like a chore… Finding new ideas became challenging as days repeated, leading me back to old God-Saeng tasks like `drinking water in the morning,' which was disappointing.''} 
P13 also realized that even the value of \textit{rest} could not be confined to an extra activity after work. Instead, it could be found in small, unplanned moments, such as a brief coffee break with a colleague, showing how flexibility allowed values to surface in unexpected ways. 
In sum, participants described consistency as evidence of diligence and progress, while also pointing to the need for flexibility when routines became repetitive or when values were discovered in unplanned moments.

\subsubsection{Authenticity vs. Performativity}
\begin{revisedenv}
Alongside the pursuit of consistency, participants navigated a persistent tension between performing God-Saeng and feeling authentic in their daily practices.

For many, the stated plan and the visible outcome of an action captured only a thin part of the work that made it possible. P4’s bite-sized mission, ``stay awake in my summer-term class,'' appeared to be a straightforward behavioral accomplishment. Yet his reflection note revealed a much richer configuration of effort: taking a 30-minute nap beforehand, stretching to reduce fatigue, arriving early to sit in the front, and actively focusing throughout the lecture. She wrote that she was proud of \textit{``all the steps I took to make this happen,''} emphasizing that the authentic meaning of her action resided not in the simple performance of staying awake, but in the layered effort behind it. By narrating these contextual details, P4 reframed performance as only the surface of a deeper lived practice.

For some participants, polished records and consistent metrics created an external appearance of the ideal God-Saeng life while obscuring internal struggles. During the debriefing interview, P9 described: 
\begin{quote}
    \textit{``I used to believe I was living my best God-Saeng, through how meticulously I managed my diary and budget tracker. But when I examined each day through the bite-sized missions, I realized how anxious and self-critical I actually was. The data looked great, but emotionally I wasn’t living the way I thought I was.''} 
\end{quote}

For many, bite-sized missions also created space to attend to the internal states that fall outside the performative frames. One of P9's value keywords was \textit{`Excitement.'} She reflected: 
\begin{quote}
    \textit{``No one has ever asked me if I had an exciting day. The world doesn’t evaluate that. Until now, I didn’t really have an outlet to think about how I could cultivate excitement in my everyday life. I realized that excitement isn’t something you perform, but something you feel in the moment, by being true to myself.''}
\end{quote}

For some participants, authenticity emerged through recognizing values already present in ordinary life. Before the bite-sized mission phase, P6 believed she lacked the three value keywords she identified, ``healing,'' ``love,' and ``reflection'', because she had never \textit{``intentionally planned nor performed executable actions to fulfill them.''} Yet in the debriefing, she mentioned, \textit{``It wasn’t that these things weren’t happening. I just hadn’t noticed them. The reflections made them visible on their own terms.''}  

Participants’ differing approaches to designing their missions further underscored this tension. Some crafted structured, outcome-oriented tasks. For example, P14 was \textit{``inspired by frameworks such as SMART goals\footnote{A framework for setting objectives that are Specific, Measurable, Achievable, Relevant, and Time-bound},''} including conditional ``if-then'' rules intended to create clear criteria for achievement. Others deliberately formulated ambiguous, qualitative missions such as ``have a good day'' or ``listen to the world.''  
\end{revisedenv}
\subsubsection{Productivity vs. Broader Values}
The emphasis on achievement as a core value was not solely a product of individual aspirations but was deeply intertwined with participants' life phases and external circumstances. Participants cited recent job placements (P7), the growth-centric startup culture (P17), academic pressures (P19), and job-seeking phases (P2) as significant drivers for prioritizing achievement, which also became a central motivation for living God-Saeng.

Yet when prompted to reflect on their values more broadly, participants highlighted relationships, rest, and other dimensions of well-being as equally or more important. P3, for example, was struck by how often photographs from her missions were labeled as meaningful because they were ``with someone,'' ``for someone,'' or ``thanks to someone,'' realizing that shared moments with others shaped her sense of well-being more than individual accomplishments. Similarly, P5, who initially valued rest, relationships, and nature, began to feel these as an added burden when framed within a productivity mindset: \textit{``I thought I had to manage not just my own life but also relationships with friends and family to live a good life. I started to feel like this was an extra layer of God-Saeng.''}

Participants also found that embedding broader values into everyday moments created meaning that productivity alone could not provide. 
\revised{For example, one of P8's missions, \textit{``take the stairs instead of waiting for the elevator,''} was originally grounded in `health' for cardiovascular exercise. But when the action was carried out, she happened to meet her mother at the entrance of their apartment building and walked up together. She wrote, \textit{``Talking with my mom as we climbed made the time pass quickly. It was nice to share that moment together.''} In the debriefing, she added that while she initially labeled the mission as growth, she later added relationships and peacefulness as more central values: \textit{``The point became not really the exercise anymore, but the time with my mom.''}} 
P23 described finding happiness on a bike ride home, appreciating the scents of plants and rain as a way of fulfilling values of sensation and joy. Likewise, P7 noted the satisfaction of expressing kindness to strangers, such as offering a seat on the subway, which allowed the value of happiness to be realized through small gestures.

These accounts illustrate how the pursuit of achievement often coexisted with, and at times conflicted with, participants’ desire to prioritize relationships, rest, and everyday joy, underscoring the ongoing negotiation between productivity-driven ideals and broader values of well-being. 

\subsection{Reinterpreting and Reshaping Well-Being}
\subsubsection{Determining Personal Value}
The underlying values identified by participants revealed a broader appreciation for social connections, rest, and everyday joy, even though their initial perceptions of God-Saeng emphasized self-improvement. In the value discovery workshop, interpersonal relationships emerged as the most prominent value, followed by achievement (success and growth, chosen by 14 participants) and rest (relaxation, ease, leisure, and healing). Other values such as art, nature, thriftiness, and joy were also mentioned, reflecting the diversity of personal priorities.

Several participants described how the workshop and daily bite-sized missions shifted their focus from achievement toward more relational or experiential values. For example, P2 noted how job-seeking pressures had previously tied their goals to performance, but through reflection they realized that \textit{``being with family without thinking about results''} was what felt most meaningful. P11 emphasized happiness as their daily mission, describing small acts (e.g., enjoying a meal with a partner) as sufficient proof of living well, contrasting with the productivity-centered image of God-Saeng.

Participants reported finding gratitude and fulfillment in integrating values into everyday surroundings.
P12 reflected on how identifying rest as a value helped them reinterpret ordinary breaks during a busy day as valuable in themselves, rather than as lost time. Participants repeatedly highlighted noticing joy in simple encounters, such as conversations with friends or small gestures of kindness, which reframed ordinary interactions as opportunities to live according to their values.

\subsubsection{Embedding Values in Everyday Action}
Participants adopted diverse strategies to implement the values they identified in everyday life through bite-sized missions.

A common strategy was to blend scheduled tasks with additional layers of personal significance. P13 described this as finding \textit{``a spoonful more''} to add to existing plans, while P10 crafted a mission around asking friends how they had been during a routine group meeting. P15 similarly noted that recognizing how ordinary tasks could fulfill personal values made them more devoted to the time spent on these tasks.

Others went beyond modification, drawing on their surroundings to design missions creatively. For example, P23, who was learning Vietnamese, set a mission to recall a word when passing by a pho restaurant, turning a casual encounter into an opportunity to reinforce learning and joy.

Some participants inverted the expected plan–act–reflect flow altogether.
P10, for instance, skipped pre-planning bite-sized missions and used the bite-sized mission card purely as an evening reflection tool. 
Each night he reviewed the day first—asking which moment felt fulfilling—and only then recorded a bite-sized mission and labeled the values it expressed. 
Through this retrospective practice, he realized that \textit{``even on tough days there was at least one moment of joy,''} a practice that gradually deepened his sense of gratitude.
P10’s account shows that values could be enacted not only through deliberate tasks but also by recognizing meaning already present in daily routines, embedding those values into life as it naturally unfolded.

Participants also discovered that values could be realized in less expected ways. P13, who initially added extra activities to pursue rest after work, came to see that a brief coffee break with a colleague also embodied rest, shifting their approach from planning additional tasks to recognizing values in everyday moments. Similarly, P19 described how reflecting at the end of the day allowed them to uncover where values had been fulfilled, even when no mission had been set in advance.

\subsubsection{Constructing Personal Well-being Narratives}
Participants engaged with the societal narrative of God-Saeng while reflecting on their own values, leading to diverse ways of constructing personal narratives of well-being. For some, this process offered clarity and validation. P10, for instance, explained, \textit{``I initially felt conflicted by the societal pressure to always strive and prove to be a better self. However, I came to validate myself that unlike others, simply finding meaning in one small thing in my day makes me feel like I’m living well.''} P11 took a similar stance by consistently setting \textit{``being happy today''} as a daily mission. While the specific actions varied—from enjoying a simple meal to spending time with a partner—this practice helped him affirm that small, everyday moments were enough to define well-being on his own terms.

Other participants, however, experienced confusion as they tried to reconcile personal values with societal expectations. P13 noted, \textit{``My bite-sized mission of spending time with friends was fulfilling, but I didn’t accomplish the productive tasks I had planned for the day. I need to figure out how to achieve both dimensions in limited amount of time.''} Likewise, P12 shared frustration when the repetition of similar missions made value fulfillment feel like an obligation rather than genuine meaning, saying it began to feel \textit{``like a chore.''} These struggles highlight how values such as friendship or rest could conflict with the productivity-oriented ideals embedded in God-Saeng, leaving participants uncertain about what it truly means to live well.

In sum, reflecting on God-Saeng alongside personal values gave some participants clarity and a sense of validation, while for others it generated confusion, underscoring the complex and sometimes contradictory intersections between societal ideals and individual meaning-making.

\section{Discussion}
Our findings show that the God-Saeng phenomenon is not just an individual pursuit but a sociotechnical practice shaped by life uncertainty, cultural ideals of diligence, and the mediating role of technology. In this section we step back from the individual cases to consider how these forces---personal, cultural, and technological---intertwine with young adults' pursuit of well-being and what this means for the design of PI systems. \revised{We first interpret God-Saeng as a socio-technical response to precarity, reflect on the tensions our participants negotiated, and then draw implications for PI technologies by questioning the taken-for-granted assumptions underlying their design.}
\begin{revisedenv}
\subsection{God-Saeng as a Socio-Technical Response to Precarity}
Young adults in our study described turning to God-Saeng routines not merely as productivity techniques but as a way to cope with persistent sense of precarity. Participants spoke about uncertainty around academic trajectories, employment, financial futures, and social expectations that felt largely outside their control. 
Looking at this phenomenon through the lens of ``lived'' situated practices ~\cite{dourish2006implications, tanenbaum2018steampunk, kaptelinin}, we may understand God-Saeng as a cultural practice through which participants navigate the existential problem of uncertainty. 

To sustain a sense of existential steadiness, participants assembled what we describe as \textit{sociotechnical infrastructures of externalized discipline.} 
Rather than relying solely on personal willpower, they configured combinations of technologies, social arrangements, and procedural rules that would ``hold them to'' their intentions when they lacked personal motivation. Through visible commitments to peers, enforced routines, or technological features that made their actions publicly or temporally accountable, they were able to anchor a series of obligations outside the ``self.'' 

In this context, \textit{planning} also played a critical role in anchoring participants' behavior to a more stable ground. In line with Suchman's account of plans as \textit{``representational resources rather than prescriptive scripts for action~\cite{suchman2007human},''} plans, here, functioned as cognitive scaffolds that made participants' intentions visible and accountable. For example, by making visible their plans and accomplishments, participants came to experience their life as more navigable, feeling more able to hold things together with and through others.
Moreover, plans offered a stable representational form onto which participants could project a desired sense of self, supporting them in finding orientation in the midst of uncertain conditions. This interpretation aligns with works in existential HCI, which highlight how people use self-constructed structures to create coherence and situate themselves amid ambiguity~\cite{cho2024reinforcing, light2017design}.

Yet these infrastructures also produced and circulated data. Timestamped photos for proof-of-action, streak counts, study hours, retrospective logs became evidence of diligence and reliability. For some participants, this datafication of their existence offered tangible signs that they were moving forward and staying disciplined. At the same time, some participants described moments when these data representations failed to match how they felt internally, amplifying anxiety, self-judgment, or comparison. This duality echoes critiques in existential HCI that warn against designs which overemphasize behavioral correctness at the expense of subjective experience~\cite{light2017design, dourish2006implications}. In our study, datafication grounded participants in stable representations of their selves and their lives, while also codifying their routines into evaluative artifacts that became an active force in their universe of meaning, shaping how they interpreted their days, their discipline, and sometimes their own worth. 

Taken together, these findings show that God-Saeng is a cultural, socio-technical coping infrastructure through which young adults negotiate precarity and reclaim agency. These practices reflect Korean norms around diligence and social accountability~\cite{kim2021mechanisms, bonk2011accountability, gelfand2004culture, kwon2011does}, yet they also illuminate broader questions within HCI about how people mobilize routines and technologies to create coherence in uncertain times.

At the same time, our study shows that within the infrastructures, tension exist between consistency and flexibility, authenticity and performativity, productivity and broader values. These tensions point toward the need for technologies that support negotiation rather than compliance. 
\end{revisedenv}
\subsection{Designing for Negotiated Tensions}
A key insight from our study is that everyday well-being is not realized through the linear pursuit of fixed goals, but through the ongoing negotiation of competing demands. Participants described the value of persistence, with repeated efforts framed as evidence of resilience and diligence. Yet when routines became rigid or repetitive, they often turned into chores rather than meaningful practices. Consistency was thus meaningful only when accompanied by flexibility, highlighting that well-being emerges from adapting routines to shifting contexts rather than adhering to them at all costs~\cite{suchman1987, rooksby2014}.

Participants often drew motivation and recognition from documenting their achievements, whether by joining \hashtag{God-Saeng challenges} on social media or using progress-tracking apps. These practices helped sustain routines, but they also turned diligence into a performance to be displayed. For some, the pursuit of self-growth masked underlying anxiety, while even values like relationships and rest were reframed as obligations such as proving one has many friends or appearing consistently extroverted. Such dynamics reveal the fragile balance between authenticity and performativity in well-being, and how technologies can amplify external validation--through approval, endorsement, and recognition--at the expense of self-fulfillment.

While achievement and productivity remained dominant, participants also articulated the importance of joy, rest, and quality relationships. Everyday moments (e.g., a bike ride, a coffee break, or a small act of kindness) emerged as equally powerful sources of meaning, yet such qualitative values are rarely foregrounded in current PI systems. These tensions demonstrate that well-being cannot be reduced to visible achievements, but is continually negotiated against broader values that often resist quantification.

Across these examples, the tensions participants navigated were not simply failures but central to how well-being was enacted. By privileging one side of these negotiations--whether persistence, performativity, or productivity--existing PI systems risk reinforcing stress and guilt. Designing for everyday well-being therefore requires a shift in focus: from driving adherence to supporting negotiation. 

\subsection{Implications for PI Systems}
\begin{revisedenv}
PI systems have long aimed to support goal-setting, behavior tracking, and reflective sense-making through structures such as progress monitoring and accountability mechanisms~\cite{li2010stage, li2011personal, epstein2020mapping}. 
Our findings show that participants' God-Saeng infrastructures enact many of these core design aspirations of PI systems: they set goals across multiple life domains, operationalize them through layered plans, sustain the practice through sociotechnical accountability features, and construct meaning from the data these systems produce.
In this sense, God-Saeng provides an unusually rich empirical site for observing how canonical PI practices take shape within a lived sociotechnical ecology rather than in controlled evaluation settings.

Yet the same participants' infrastructures also question the dominant assumptions that underlie PI systems.
Participants experienced these tools as simultaneously stabilizing and pressurizing, supportive in some moments yet anxiety-inducing in others. The usefulness of data, routines, and accountability mechanisms depended not only on behavioral outcomes but on participants’ emotional states, cultural expectations, and existential orientations.
This suggests that the central question for PI design is not simply \textit{how} to motivate adherence, optimize routines, or improve behavioral outcomes, but rather \textit{when, how,} and \textit{for whom} PI technologies become supportive, burdensome, or reinterpreted in everyday life.

This resonates with recent PI and existential HCI scholarship showing that the role of PI systems cannot be fully understood without considering users’ broader existential orientations~\cite{rapp2019behav, rutjes, ryan2022, rapp2023exploring}.  
Our findings do not imply that existing design goals of PI (e.g., tracking, feedback, adherence) should be discarded. Instead, they foreground the \textit{reciprocal nature of PI use.} 
As we have observed, individuals appropriate PI tools to meet their own situated needs, and the systems, in turn, participate in shaping how users understand their progress, emotional states, and even their sense of self. 

Because of this reciprocity, whereby PI tools participate in users' meaning-making, they cannot be solely understood or designed as serving a predefined designer-intended purpose. Instead, designers should be responsible for the ways these systems become entangled with users' broader emotional, cultural, and existential life. This suggests a need for PI systems to also provide a space where users can reflexively examine how a tool is influencing them as a whole, as culturally and socially situated beings dealing with existential matters, and not only in relation to the specific goal (such as maintaining routines consistently) for which they initially adopted it. 

The following implications build on this orientation. We outline design directions for PI systems that can help users enable meaning-making beyond data produced, foreground values over metrics, and cultivate awareness of how technologies shape their lived experience in a broader cultural context. 

\end{revisedenv}
\subsubsection{Enabling Meaning-Making Beyond Data}
Current PI systems often constrain reflection to the data they collect~\cite{cho2022reflection, lee2025designing, epstein2016beyond, li2012using, howe2022design}. System-driven supports such as automatic feedback or predefined insights, while efficient, can narrow reflection and operate as another form of persuasive technology~\cite{li2012using, cho2022reflection}. User-driven approaches, though more engaging, tend to channel interpretations through data representations that have already been preprocessed by the system~\cite{toebosch2024non, cho2022reflection}. As a result, opportunities for meaning-making are frequently reduced to singular pathways—most often, progress toward goals.

Our findings showed how participants went beyond such singular framings by layering multiple values onto mundane actions. For example, a simple act such as offering a seat to a stranger on the subway was initially not tied to any explicit goal, but it was later described as a source of happiness and connection. Across such accounts, the same everyday action could hold multiple meanings: at times framed as discipline or self-care, at others as relational, emotional, or environmental. This multiplicity shows that meaning-making cannot be captured by a single metric or goal but emerges from diverse reinterpretations of lived experience.

Designing PI systems to better support such diversity requires moving beyond metrics of progress. One direction is to scaffold reflection by prompting users to surface contextual cues~\cite{jung2025counterstress}, such as who an activity was shared with, what emotions were involved, or what environmental conditions shaped the experience, alongside goal-related information. Importantly, such cues should not be treated merely as data points but as materials through which richer meaning can be constructed.  

Another direction is to open interpretive space by offering diverse thematic perspectives through which users might reinterpret their experiences. For example, Song et al.~\cite{song2025exploreself} demonstrated how narrative prompts and Socratic questions allowed users to build storylines around their challenges, granting them agency in making sense of their experiences. Extending this approach, PI systems could integrate narrative scaffolds with collected data and contextual cues, providing users with multiple ``doors'' into reflection. In doing so, systems can better support meaning-making that acknowledges well-being as a constructed practice, rather than reducing it to progress toward predefined goals. 

\subsubsection{Supporting Value Negotiation}

Prior PI research has emphasized goal evolution—helping people adjust or switch goals as their context or life phase changes (e.g., illness, new job, semester transitions)~\cite{li2011personal,epstein2015lived, ekhtiar2025changing, epstein2020mapping, salmela2000women}.
Our findings, however, reveal that \emph{value negotiation is continuous and multi-dimensional}. Participants did not only revise goals occasionally but constantly balanced and re-weighted priorities such as productivity and rest or authenticity and social recognition in everyday micro-moments. This ongoing negotiation was not an exceptional adjustment but a defining feature of how well-being was practiced day to day.

Building on this insight, PI systems should foreground tensions among values rather than conceal them behind single metrics of achievement or failure. Instead of reporting whether a numeric goal was met, systems could highlight which values were fulfilled and how trade-offs played out. For example, a late-night study session might advance achievement while diminishing rest. By surfacing such frictions as expected features of everyday well-being, systems can help users treat negotiation itself as a normal and constructive practice rather than a sign of lapse. 

One way to operationalize this principle is through value-oriented perspectives layered on top of existing data. A \emph{value-lens extension} could allow users to view their tracked activities through different lenses (e.g., health, connection, creativity) and replay a day or week by toggling among them. Overlapping highlights would reveal where activities nourished several values at once or where prioritizing one value compromised another. Rather than prescribing normative actions, the system would foreground these overlaps and frictions, inviting users to annotate tensions and, if desired, re-balance priorities going forward.

By making tensions visible and normalizing their ongoing negotiation, PI systems can move beyond episodic goal changes to support the lived, continuous balancing that characterizes everyday well-being.

\subsubsection{Designing with Sensitivity to Cultural Contexts}

The pursuit of well-being is never universal but reflects the values and narratives that a society promotes--whether spiritual fulfillment, economic productivity, or aesthetic self-care~\cite{compton2001values}.
Our findings highlight how such cultural forces shape both the motivation to pursue well-being and the technological channels through which it is expressed.  
In the South Korean context, diligence and perseverance are not merely personal preferences but moral expectations tied to social worth~\cite{kee2008influences,kohls2001learning}.  
South Korea’s collectivist orientation amplifies sensitivity to external evaluation~\cite{eckhardt2002culture,tan2015role}, while commercial wellness platforms further reinforce this dynamic by rewarding public displays of discipline and achievement~\cite{seberger2024better,de2022outlining}.  
Technology thus does more than support well-being practices--it co-produces local ideals of what ``living well'' means.

For PI design, this calls for cultural attunement at multiple levels. Prior works have underscored the risks of one-size-fits-all metrics, such as in menstrual or reproductive health tracking, where systems must accommodate diverse religious and cultural norms~\cite{ibrahim2024islamically, ibrahim2024tracking, ibrahim2024expanding, song2024typing}. Our study extends this point by showing the need to examine the broader forces that actively shape local ideals in the first place. Rather than only tailoring interfaces to stated community needs, designers can adopt a more reflexive stance, considering how platform logics and social expectations construct the very goals that people pursue~\cite{sengers2005reflective, cho2019toward}. Supporting users in recognizing and, when desired, questioning these shaping aspects allows PI systems to move from simple cultural adaptation toward a more nuanced engagement with evolving sociotechnical conditions.

Finally, culture is dynamic~\cite{schein1983organizational, erez2004dynamic, lotman2013dynamics}. Just as God-Saeng itself emerged from shifting economic conditions, digital trends, and youth anxieties, designers must treat cultural sensitivity as an ongoing practice rather than a one-time localization step. Continuous engagement with communities and iterative, participatory research can help PI systems remain responsive to the changing cultural landscapes in which everyday well-being is defined.

\section{Limitations and Future Work}
\begin{revisedenv}
    Our study offers a situated account of how young adults in South Korea appropriate sociotechnical routines to navigate precarity, yet several limitations also open avenues for future research. First, our analysis is grounded in a specific cultural moment and demographic. While this situatedness is central to our methodological orientation, it necessarily narrows the empirical scope. 
    
    \finalrevised{Participants were recruited through purposive sampling, and engagement with God-Saeng was defined through self-described positioning rather than a standardized or validated scale. While this choice enabled us to select participants through their lived interpretations and relationships to the phenomenon, privileging participants' first-person perspectives, it may limit internal validity and consistency of the participant pool.}

    \finalrevised{Our workshop and interview design relied on pre-existing peer groups to foster comfort and shared reflection. While familiar peers can support richer recall and articulation of culturally shared practices, they may also introduce social pressures, such as aligning their answers with the group's shared norms. Future work could explore alternative group compositions to surface different dynamics.}

    Related lifestyle movements in other regions, for example, the \textit{That Girl} Challenge~\cite{CBC2021ThatGirl}, a viral social media trend encouraging individuals to become ``the best version of themselves'' appear to grapple with comparable tensions between productivity, authenticity, and stability. \finalrevised{However, we did not perform a comparative analysis with these supposedly similar phenomena, so we cannot identify what aspects of God-Saeng can be generalized across different youth self-improvement practices.} Future comparative inquiries could examine how such practices differ or converge across contexts, offering a more multi-sited understanding of how sociotechnical routines become coping resources under contemporary uncertainty.

    While this work focuses on articulating the implications for PI systems, future research could prototype and evaluate higher-fidelity systems that embody these principles. Such work would extend our contribution by examining how systems might scaffold not only behavioral tracking but also the interpretive work through which users navigate sociotechnical tensions.

    Finally, our 10-day mission phase offered a window into participants’ everyday meaning-making, but longer-term engagements may reveal additional dynamics. Precarity and personal routines often shift over seasons, life transitions, and evolving circumstances~\cite{rapp2023exploring, rapp2022}. Longitudinal fieldwork over several months could show how practices stabilize or transform over time, and how sociotechnical infrastructures adapt alongside users’ changing conditions.
\end{revisedenv}

\section{Conclusion}
\revised{This paper examined God-Saeng as a culturally situated, socio-technical practice through which young adults in South Korea navigate contemporary uncertainty. We showed how God-Saeng functions as a coping infrastructure, a way of reclaiming coherence, control, and purposefulness amid ongoing precarity. In doing so, we extend HCI's growing attention to existential concerns, illustrating how technologies become meaningful not only through what they track or optimize but through how they scaffold people’s efforts to orient themselves in their lived worlds.}

Our findings emphasize that everyday well-being is a constructed and negotiated practice rather than a linear pursuit of predefined goals. \revised{Moreover, the study findings highlight that while God-Saeng infrastructures enact many dominant PI aspirations, yet they also exemplify how these aspirations can become burdensome when detached from the existential, emotional, and cultural conditions of people’s lives. This suggests a need for PI systems that help users interpret, not merely perform, by opening reflexive spaces, foregrounding values over metrics, and supporting the active negotiation of meaning within one's dynamic sociotechnical and cultural ecologies.}

Beyond the South Korean context, our findings suggest \revised{the importance of culturally and existentially aware PI systems that not only document behavior but also open room for users to understand how personal aspirations intersect with societal pressures, uncertainty, and everyday forms of care}.

\begin{acks}
We thank our participants in the study for their time and efforts. We also thank KAIST DxD lab members for their feedback on the study design, and especially Hyunseung Lim for his feedback on the early version of our draft. This work was supported through the KAIST Undergraduate Research Program. 
\end{acks}

\bibliographystyle{ACM-Reference-Format}
\bibliography{biblography}

\end{CJK}
\end{document}